# **Kitsune**: An Ensemble of Autoencoders for Online Network Intrusion Detection

Yisroel Mirsky, Tomer Doitshman, Yuval Elovici and Asaf Shabtai
Ben-Gurion University of the Negev
{yisroel, tomerdoi}@post.bgu.ac.il, {elovici, shabtaia}@bgu.ac.il

***Abstract*—Neural networks have become an increasingly popular solution for network intrusion detection systems (NIDS). Their capability of learning complex patterns and behaviors make them a suitable solution for differentiating between normal traffic and network attacks. However, a drawback of neural networks is the amount of resources needed to train them. Many network gateways and routers devices, which could potentially host an NIDS, simply do not have the memory or processing power to train and sometimes even execute such models. More importantly, the existing neural network solutions are trained in a supervised manner. Meaning that an expert must label the network traffic and update the model manually from time to time.**

**In this paper, we present Kitsune: a plug and play NIDS which can learn to detect attacks on the local network, without supervision, and in an efficient online manner. Kitsune's core algorithm (*KitNET*) uses an ensemble of neural networks called autoencoders to collectively differentiate between normal and abnormal traffic patterns. *KitNET* is supported by a feature extraction framework which efficiently tracks the patterns of every network channel. Our evaluations show that Kitsune can detect various attacks with a performance comparable to offline anomaly detectors, even on a Raspberry PI. This demonstrates that Kitsune can be a practical and economic NIDS.**

## I. Introduction

The number of attacks on computer networks has been increasing over the years [1]. A common security system used to secure networks is a network intrusion detection system (NIDS). An NIDS is a device or software which monitors all traffic passing a strategic point for malicious activities. When such an activity is detected, an alert is generated, and sent to the administrator. Conventionally an NIDS is deployed at a single point, for example, at the Internet gateway. This *point* deployment strategy can detect malicious traffic entering and leaving the network, but not malicious traffic traversing the network itself. To resolve this issue, a *distributed* deployment strategy can be used, where a number of NIDSs are be connected to a set of strategic routers and gateways within the network.

Over the last decade many machine learning techniques have been proposed to improve detection performance [2], [3], [4]. One popular approach is to use an artificial neural network (ANN) to perform the network traffic inspection. The benefit of using an ANN is that ANNs are good at learning complex non-linear concepts in the input data. This gives ANNs a great advantage in detection performance with respect to other machine learning algorithms [5], [2].

The prevalent approach to using an ANN as an NIDS is to train it to classify network traffic as being either normal or some class of attack [6], [7], [8]. The following shows the typical approach to using an ANN-based classifier in a *point* deployment strategy:

1) Have an expert collect a dataset containing both normal traffic and network attacks.
2) Train the ANN to classify the difference between *normal* and *attack* traffic, using a strong CPU or GPU.
3) Transfer a copy of the trained model to the network/organization's NIDS.
4) Have the NIDS execute the trained model on the observed network traffic.

In general, a *distributed* deployment strategy is only practical if the number of NIDSs can economically scale according to the size of the network. One approach to achieve this goal is to embed the NIDSs directly into inexpensive routers (i.e., with simple hardware). We argue that it is impractical to use ANN-based classifiers with this approach for several reasons:

**Offline Processing.** In order to train a supervised model, all labeled instances must be available locally. This is infeasible on a simple network gateway since a single hour of traffic may contain millions of packets. Some works propose offloading the data to a remote server for model training [9] [3]. However, this solution may incur significant network overhead, and does not scale.

**Supervised Learning.** The labeling process takes time and is expensive. More importantly, what is considered to be normal depends on the local traffic observed by the NIDS. Furthermore, in attacks change overtime and while new ones are constantly being discovered [10], so continuous maintainable of a malicious attack traffic repository may be impractical. Finally, classification is a closed-world approach to identifying concepts. In other words, a classifier is trained to identify the classes provided in the training set. However, it is unreasonable to assume that all possible classes of *malicious* traffic can be collected and placed in the training data.

**High Complexity.** The computational complexity of an ANN

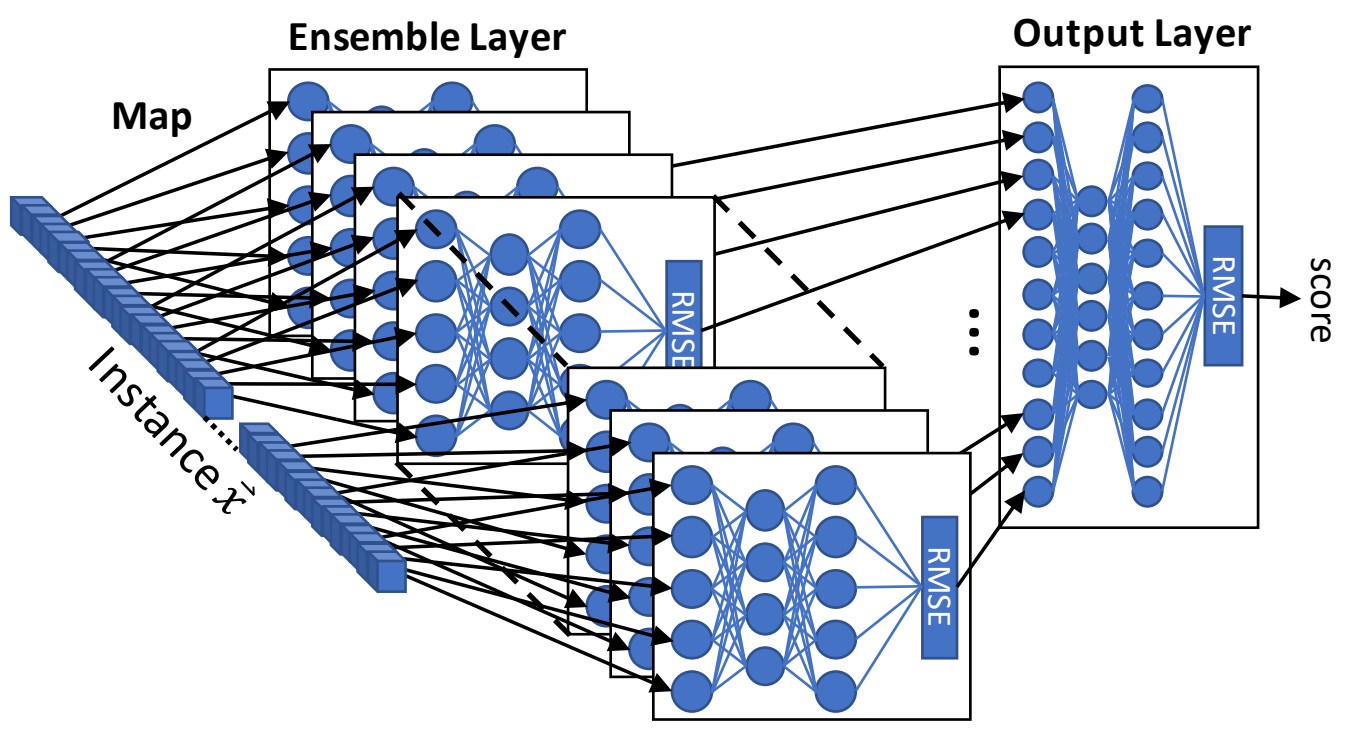

Fig. 1: An illustration of **Kitsune**'s anomaly detection algorithm *KitNET*.

grows exponentially with number of neurons [11]. This means that an ANN which is deployed on a simple network gateway, is restricted in terms of its architecture and number of input features which it can use. This is especially problematic on gateways which handle high velocity traffic.

In light of the challenges listed above, we suggest that the development of an ANN-based network intrusion detector, which is to be deployed and trained on routers in a distributed manner, should adhere to the following restrictions:

**Online Processing.** After the training or executing the model with an instance, the instance is immediately discarded. In practice, a small number of instances can be stored at any given time, as done in stream clustering [12].

**Unsupervised Learning.** Labels, which indicate explicitly whether a packet is *malicious* or *benign*, are not used in the training process. Other meta information can be used so long as acquiring the information does not delay the process.

**Low Complexity.** The packet processing rate must exceed the expected maximum packet arrival rate. In other words, we must ensure that there is no queue of packets awaiting to be processed by the model.

In this paper, we present **Kitsune**: a novel ANN-based NIDS which is online, unsupervised, and efficient. A Kitsune, in Japanese folklore, is a mythical fox-like creature that has a number of tails, can mimic different forms, and whose strength increases with experience. Similarly, **Kitsune** has an ensemble of small neural networks (autoencoders), which are trained to mimic (reconstruct) network traffic patterns, and whose performance incrementally improves overtime.

The architecture of **Kitsune's** anomaly detection algorithm (*KitNET*) is illustrated in Fig. 1. First, the features of an instance are mapped to the visible neurons of the ensemble. Next, each autoencoder attempts to reconstruct the instance's features, and computes the reconstruction error in terms of root mean squared errors (RMSE). Finally, the RMSEs are forwarded to an output autoencoder, which acts as a non-linear voting mechanism for the ensemble. We note that while training **Kitsune**, no more than one instance is stored in memory at a time. *KitNET* has one main parameter, which is the maximum number of inputs for any given autoencoder in the ensemble. This parameter is used to increase the algorithm's speed with a modest trade off in detection performance.

The reason we use autoencoders is because (1) they can trained in an unsupervised manner, and (2) they can be used for anomaly detection in the event of a poor reconstruction. The reason we propose using an ensemble of small autoencoders, is because they are more efficient and can be less noisier than a single autoencoder over the same feature space. From our experiments, we found that **Kitsune** can increase the packet processing rate by a factor of five, and provide a detection performance which rivals other an offline (batch) anomaly detectors.

In summary, the contributions of this paper as follows:

- A novel autoencoder-based NIDS for simple network devices (**Kitsune**), which is lightweight and plug-and-play. To the best of our knowledge, we are the first to propose the use of autoencoders with or without ensembles for online anomaly detection in computer networks. We also present the core algorithm (*KitNET*) as a generic online unsupervised anomaly detection algorithm, and provide the source code for download.[1]
- A feature extraction framework for dynamically maintaining and extracting implicit contextual features from network traffic. The framework has a small memory footprint since the statistics are updated incrementally over damped windows.
- An online technique for automatically constructing the ensemble of autoencoders (i.e., mapping features to ANN inputs) in an unsupervised manner. The method involves the incremental hierarchal clustering of the feature-space (transpose of the unbounded dataset), and bounding of cluster sizes.
- Experimental results on an operational IP camera video surveillance network, IoT network, and a wide variety of attacks. We also demonstrate the algorithm's efficiency, and ability to run on a simple router, by performing benchmarks on a Raspberry PI.

The rest of the paper is organized as follows: Section II discusses related work in the domain of online anomaly detection. Section III provide a background on autoencoders and how they work. Section IV presents **Kitsune**'s framework and it's entire machine learning pipeline. Section V presents experimental results in terms of detection performance and run-time performance. Finally, in section VII we present our conclusion.

## II. Related Work

The domain of using machine learning (specifically anomaly detection) for implementing NIDSs was extensively researched in the past [13], [14], [15], [16], [17]. However, these solutions usually do not have any assumption on the resources of the machine running training or executing the model, and therefore are either too expensive to train and execute on simple gateways, or require a labeled dataset to perform the training process.

Several previous works have proposed online anomaly detection mechanisms using different lightweight algorithms. For example, the PAYL IDS which models simple histograms of packet content [18] or the kNN algorithm [19]. These methods are either very simple and therefore produce very poor results, or require accumulating data for the training or detection.

[1] The source code for *KitNET* is available for download at: https://github.com/ymirsky/KitNET-py.

A popular algorithm for network intrusion detection is the ANN. This is because of its ability to learn complex concepts, as well as the concepts from the domain of network communication [17]. In [20], the authors evaluated the ANN, among other classification algorithms, in the task of network intrusion detection, and proposed a solution based on an ensemble of classifiers using connection-based features. In [8], the authors presented a modification to the back propagation algorithm to increase the speed of an ANN's training process. In [7], the authors used multiple ANN-based classifiers, where each one was trained to detect a specific type of attack. In [9], the authors proposed a hierarchal method where each packet first passes through an anomaly detection model, then if an anomaly is raised, the packet is evaluated by a set of ANN classifiers where each classifier is trained to detect a specific attack type.

All of the aforementioned papers which use ANNs, are either supervised, or are not suitable for a simple network gateway. In addition, some of the works assume that the training data can be stored and accumulated which is not the case for simple network gateways. Our solution enables a plug-and-play deployment which can operate at much faster speeds than the aforementioned models.

With regards to the use of autoencoders: In [21], the authors used an ensemble of deep neural networks to address object tracking in the online setting. Their proposed method uses a stacked denoising autoencoder (SDAE). Each layer of the SDAE serves as a different feature space for the raw image data. The scheme transforms each layer of the SDAE to a deep neural network which is used as discriminative binary classifier. Although the authors apply autoencoders in an online setting, they did not perform anomaly detection, nor address the challenge of real-time processing (which is great challenge with deep neural networks). Furthermore, training a deep neural network is complex and cannot be practically performed on a simple network device. In [22] and [23], the authors propose the use of autoencoders to extract features from datasets in order to improve the detection of cyber threats. However, the autoencoders themselves were not used for anomaly detection. Ultimately, the authors use classifiers to detect the cyber threats. Therefore, their solution requires an expert to label instances, whereas our solution is unsupervised, and plug-and-play.

In [24], the authors proposed the generic use of an autoencoder for detecting anomalies. In [19], the authors use autoencoders to detect anomalies in power grids. These works differ from ours because (1) they are not online, (2) the architecture used by the authors is not lightweight and scalable as an ensemble, and (3) has not been applied to network intrusion detection. We note that part of this paper's contribution is an appropriate feature extraction framework, which enables the use of autoencoders in the online network setting.

## III. Background: Autoencoders

Autoencoders are the foundation building blocks of **Kitsune**. In this section we provide a brief introduction to autoencoders; what they are, and how they work. To describe the training and execution of an auto encoder we will refer to the example in Fig. 2.

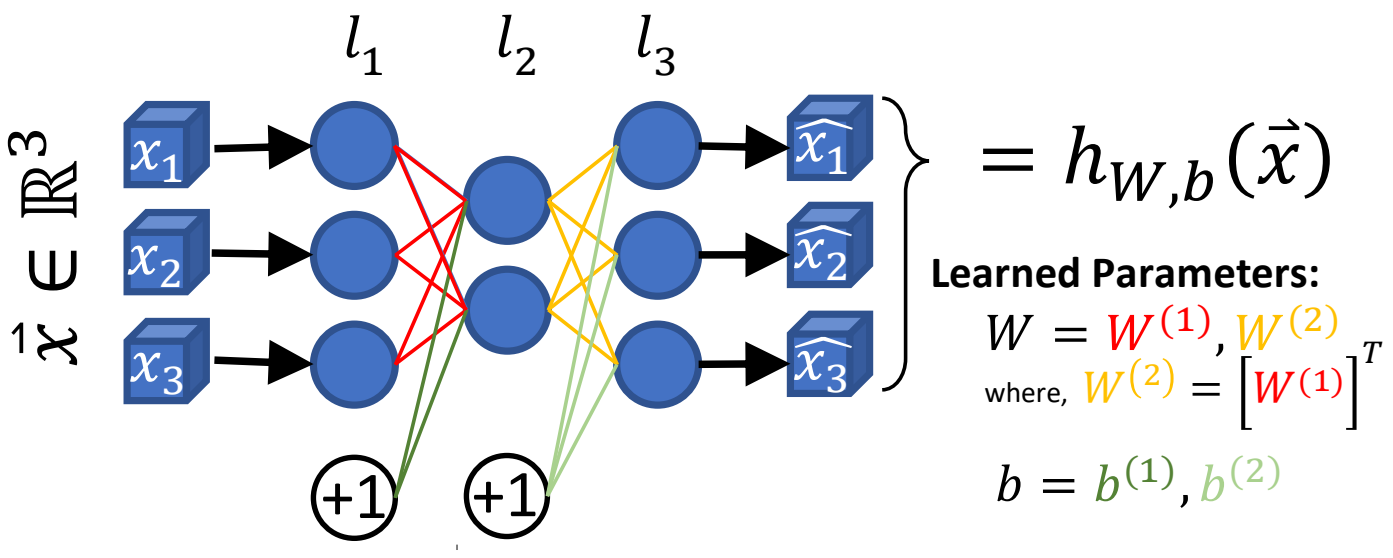

Fig. 2: An example autoencoder with one compression layer, which reconstructs instances with three features.

### A. Artificial Neural Networks

ANNs are made up of layers of neurons, where each layer is connected sequentially via synapses. The synapses have associated weights which collectively define the concepts learned by the model. Concretely, let $l^{(i)}$ denote the $i$-th layer in the ANN, and let $\|l^{(i)}\|$ denote the number of neurons in $l^{(i)}$. Finally, let the total number of layers in the ANN be denoted as $L$. The weights which connect $l^{(i)}$ to $l^{(i+1)}$ are denoted as the $\|l^{(i)}\|$-by-$\|l^{(i+1)}\|$ matrix $W^{(i)}$ and $\|l^{(i+1)}\|$ dimensional bias vector $\vec{b}^{(i)}$. Finally, we denote the collection of all parameters $\theta$ as the tuple $\theta \equiv (W, b)$, where $W$ and $b$ are the weights of each layer respectively. Fig 2 illustrates how the weights form the synapses of each layer in an ANN.

There are two kinds of layers in an ANN: visible layers and hidden layers. The visible layer receives the input instance $\vec{x}$ with an additional bias variable (a constant value of 1). $\vec{x}$ is a vector of numerical features which describes the instance, and is typically normalized to fall out approximately on the range of $[-1, +1]$ (e.g., using 0-1 normalization or zscore normalization)[25]. The difference between the visible layer and the hidden layers is that the visible layer is considered to be precomputed, and ready to be passed to the second layer.

### B. Executing an ANN

To execute an ANN, $l^{(2)}$ is activated with the output of $l^{(1)}$ (i.e., $\vec{x}$) weighted with $W^{(1)}$, then the output $l^{(2)}$ weighted with $W^{(2)}$ is used to activate $l^{(3)}$, and so on until the final layer has been activated. This process is known as *forward-propagation*. Let $\vec{a}^{(i)}$ be the $\|l^{(i)}\|$ vector of outputs from the neurons in $l^{(i)}$. To obtain $a^{(i+1)}$, we pass $\vec{a}^{(i)}$ through $l^{(i+1)}$ by computing

$$a^{(i+1)} = f\left(W^{(i)} \cdot \vec{a}^{(i)} + \vec{b^{(i)}}\right) \quad (1)$$

where $f$ is what's known as the neuron's activation function. A common activation function, and what we use in **Kitsune**, is the sigmoid function, defined as

$$f(\vec{x}) = \frac{1}{1 + e^{\vec{x}}} \quad (2)$$

**Algorithm 1:** The *back-propagation* algorithm for performing batch-training of an ANN.

**procedure:** $\texttt{train}_{GD}(\theta, X, Y, max_iter)$
$\theta \leftarrow \mathcal{U}(-\frac{1}{\|l^{(1)}\|}, \frac{1}{\|l^{(1)}\|})$. ▷ random initialization
cur_iter ← 0
**while** *cur_iter* ≤ *max_iter* **do**
  $A, Y' \leftarrow h_\theta(X)$ ▷ *forward propagation*
  $deltas \leftarrow b_\theta(Y, Y')$ ▷ *backward propagation*
  $\theta \leftarrow \texttt{GD}_\ell(A, deltas)$ ▷ *weight update*
  cur_iter++
**end**
return $\theta$

Finally, we define output of the last layer to be denoted as $\vec{y'} = a^{(L)}$. Let the function $h$ be the full layer-wise forward-propagation from the input $\vec{x}$ until the final output $\vec{y'}$, denoted

$$h_\theta(\vec{x}) = \vec{y'} \tag{3}$$

### C. Training an ANN

The output of an ANN depends on the training set, and the training algorithm. A training dataset is composed of instances $\vec{x} \in X$ and the respective expected outputs $\vec{y} \in Y$ (e.g., the labels in classification). During training, the weights of the ANN are tuned so that $h_\theta(\vec{x}) = \vec{y}$. A common algorithm for training an ANN (i.e., finding the optimum $W$ and $\vec{b}$ given $X, Y$) is known as the *back-propagation* algorithm [6].

The *back-propagation* algorithm involves a step where the error between the predicted $y'$ and the expected $y$ is propagated from the output to each neuron. During this process, the errors between a neuron's actual and expected activations are stored. We denote this backward-propagation step as the function $b_\theta(Y, Y')$. Given some execution of $h_\theta$, let $A$ be every neuron's activation and let $Delta$ be the activation errors of all neurons.

Given the current $A$ and $Delta$, and a set learning rate $\ell \in (0, 1]$, we can incrementally optimize $W$ and $b$ by performing the Gradient Descent (GD) algorithm [26]. All together, the *back-propagation* algorithm is as follows:

The above process is referred to as batch-training. This is because for each iteration, GD updates the weights according to the collective errors of all instances in $X$. Another approach is stochastic-training, where Stochastic Gradient Descent (SGD) is used instead of GD. In SGD, the weights are updated according to the errors of each instance individually. With SGD, the *back propagation* algorithm becomes:

The difference between GD and SGD is that GD converges on an optimum better than SGD, but SGD initially converges faster [27]. For a more elaborate explanation of the *back-propagation* algorithm, we refer the reader to [26].

In **Kitsune**, we use SGD with a max_iter of 1. In other words, while we are in the training phase, if a new instance arrives we perform a single iteration of the inner loop in Algorithm 2, discard the instance, and then wait for the next. This way we learn only once from each observed instance, and remain an online algorithm.

**Algorithm 2:** The *back-propagation* algorithm for performing stochastic-training of an ANN.

**procedure:** $\texttt{train}_{SGD}(\theta, X, Y, max_iter)$
$\theta \leftarrow \mathcal{U}(-\frac{1}{\|l^{(1)}\|}, \frac{1}{\|l^{(1)}\|})$. ▷ random initialization
cur_iter ← 0
**while** *cur_iter* ≤ *max_iter* **do**
  **for** $x_i$ *in* $\|X\|$ **do**
    $A, y' \leftarrow h_\theta(x_i)$ ▷ *forward propagation*
    $deltas \leftarrow b_\theta(\vec{y}, \vec{y'})$ ▷ *backward propagation*
    $\theta \leftarrow \texttt{GD}_\ell(A, deltas)$ ▷ *weight update*
  **end**
  cur_iter++
**end**
return $\theta$

### D. Autoencoders

An autoencoder is an artificial neural network which is trained to reconstruct it's inputs (i.e., $X = Y$). Concretely, during training, an autoencoder tries to learn the function

$$h_\theta(\vec{x}) \approx \vec{x} \tag{4}$$

It can be seen that an autoencoder is essentially trying to learn the identity function of the original data distribution. Therefore, constraints are placed on the network, forcing it to learn more meaningful concepts and relationships between the features in $\vec{x}$. The most common constraint is to limit the number of neurons in the inner layers of the network. The narrow passage causes the network to learn compact encodings and decodings of the input instances.

As an example, Fig. 2 illustrates an autoencoder which receives an instance $\vec{x} \in \mathbb{R}^3$ at layer $l^{(1)}$, (2) encodes (compresses) $\vec{x}$ at layer $l^{(2)}$, and (3) decodes the compressed representation of $\vec{x}$ at layer $L^{(3)}$. If an autoencoder is symmetric in layer sizes, the same (mirrored) weights can be used for coding and decoding [28]. This trick reduces the number of calculations needed during training. For example, in Fig 2, $W^{(2)} = [W^{(1)}]^T$.

### E. Anomaly Detection with Autoencoders

Autoencoders have been used for many different machine learning tasks. For example, generating new content [29], and filtering out noise from images [30]. In this paper, we are interested in using autoencoders for anomaly detection.

In general, an autoencoder trained on $X$ gains the capability to reconstruct unseen instances from the same data distribution as $X$. If an instance does not belong to the concepts learned from $X$, then we expect the reconstruction to have a high error. The reconstruction error of the instance $\vec{x}$ for a given autoencoder, can be computed by taking the root mean squared error (RMSE) between $\vec{x}$ and the reconstructed output $\vec{y'}$. The RMSE between two vectors is defined as

$$\texttt{RMSE}(\vec{x}, \vec{y}) = \sqrt{\frac{\sum_{i=1}^{n}(x_i - y_i)^2}{n}} \tag{5}$$

where $n$ is the dimensionality of the input vectors.

Let $\phi$ be the anomaly threshold, with an initial value of $-1$, and let $\beta \in [1, \infty)$ be some given sensitivity parameter. One can apply an autoencoder to the task of anomaly detection by performing the following steps:

1) **Training Phase:** Train an autoencoder on clean (normal) data. For each instance $x_i$ in the training set $X$:
   a) Execute: $s = \texttt{RMSE}\left(\vec{x}, h_\theta(\vec{x})\right)$
   b) Update: if($s \geq \phi$) then $\phi \leftarrow s$
   c) Train: Update $\theta$ by learning from $x_i$
2) **Execution Phase:**
   When an unseen instance $\vec{x}$ arrives:
   a) Execute: $s = \texttt{RMSE}\left(\vec{x}, h_\theta(\vec{x})\right)$
   b) Verdict: if($s \geq \phi\beta$) then *Alert*

The process in which **Kitsune** performs anomaly detection over an ensemble of autoencoders will be detailed later in section IV-D

### F. Complexity

In order to activate the layer $l^{(i+1)}$, one must perform the matrix multiplication $W^{(i)} \cdot \vec{a}^{(i)}$ as described in (1). Therefore, the complexity of activating layer $l^{(i+1)}$ is $O\left(l^{(i)} \cdot l^{(i+1)}\right)$.[2] Therefore, the total complexity of executing an ANN is dependent on the number of layers, and the number of neurons in each layer. The complexity of training an ANN on a single instance using SDG (Algorithm 2) is roughly double the complexity of execution. This is because of the backward-propagation step.

We note that autoencoders can be deep (have many hidden layers). In general, deeper and wider networks can learn concepts which are more complex. However, as shown above, deep networks can be computationally expensive to train and execute. This is why in *KitNET* we ensure that each autoencoder is limited to three layers with at most seven visible neurons.

## IV. THE KITSUNE NIDS

In this section we present the **Kitsune** NIDS: the packet preprocessing framework, the feature extractor, and the core anomaly detection algorithm. We also discuss the complexity of the anomaly detection algorithm and provide a bound on its runtime performance.

### A. Overview

**Kitsune** is a *plug-and-play* NIDS, based on neural networks, and designed for the efficient detection of abnormal patterns in network traffic. It operates by (1) monitoring the statistical patterns of recent network traffic, and (2) detecting anomalous patterns via an ensemble of autoencoders. Each autoencoder in the ensemble is responsible for detecting anomalies relating to a specific aspect of the network's behavior. Since **Kitsune** is designed to run on simple network routers, and in real-time, **Kitsune** has been designed with small memory footprint and a low computational complexity.

[2]The modern approach is to accelerate these operations using a GPU. However, in this paper, we assume that no GPU is available. This is the case for a simple network router.

**Kitsune**'s framework is composed of the following components:

- **Packet Capturer:** The external library responsible for acquiring the raw packet. Example libraries: NFQueue[31], afpacket[32], and tshark[33] (Wireshark's API).
- **Packet Parser:** The external library responsible for parsing raw packets to obtain the meta information required by the Feature Extractor. Example libraries: Packet++[3], and tshark.
- **Feature Extractor (FE):** The component responsible for extracting $n$ features from the arriving packets to create creating the instance $\vec{x} \in \mathbb{R}^n$. The features of $\vec{x}$ describe the a packet, and the network channel from which it came.
- **Feature Mapper (FM):** The component responsible for creating a set of smaller instances (denoted as $\mathbf{v}$) from $\vec{x}$, and passing $\mathbf{v}$ to the in the Anomaly Detector (AD). This component is also responsible for learning the mapping, from $\vec{x}$ to $\mathbf{v}$.
- **Anomaly Detector (AD):** The component responsible for detecting abnormal packets, given a packet's representation $\mathbf{v}$.

Since the Packet Capturer and Packet Extractor are not the contributions of this paper, we will focus on the FE, FM, and AD components. We note that FM and AD components are task generic (i.e., solely depend on the input features), and therefore can be reapplied as a generic online anomaly detection algorithm. Moreover, we refer to the generic algorithm in the AD component as *KitNET*.

*KitNET* has one main input parameter, $m$: the maximum number of inputs for each autoencoder in *KitNET*'s ensemble. This parameter affects the complexity of the ensemble in *KitNET*. Since $m$ involves a trade-off between detection and runtime performance, the user of **Kitsune** must decide what is more important (detection rate vs packet processing rate). This trade-off is further discussed later in section V.

The FM and AD have two modes of operation: *train-mode* and *exec-mode*. For both components, *train-mode* transitions into *exec-mode* after some user defined time limit. A component in *train-mode* updates its internal variables with the given inputs, but does not generate outputs. Conversely, a component in *exec-mode* does not update its variables, but does produce outputs.

In order to better understand how **Kitsune** works, we will now describe the process which occurs when a packet acquired, as depicted in Fig. 3:

1) The **Packet Capturer** acquires a new packet and passes the raw binary to the Packer Parser.
2) The **Packet Parser** receives the raw binary, parses the packet, and sends the meta information of the packet to the FE. For example, the packet's arrival time, size, and network addresses.
3) The **FE** receives this information, and uses it to retrieve over 100 statistics which are used to implicitly describe the current state of the channel from which the packet came. These statistics form the instance $\vec{x} \in \mathbb{R}^n$, which is passed to the FM.

[3]The Packet++ project can be found on GitHub: `https://github.com/seladb/PcapPlusPlus`

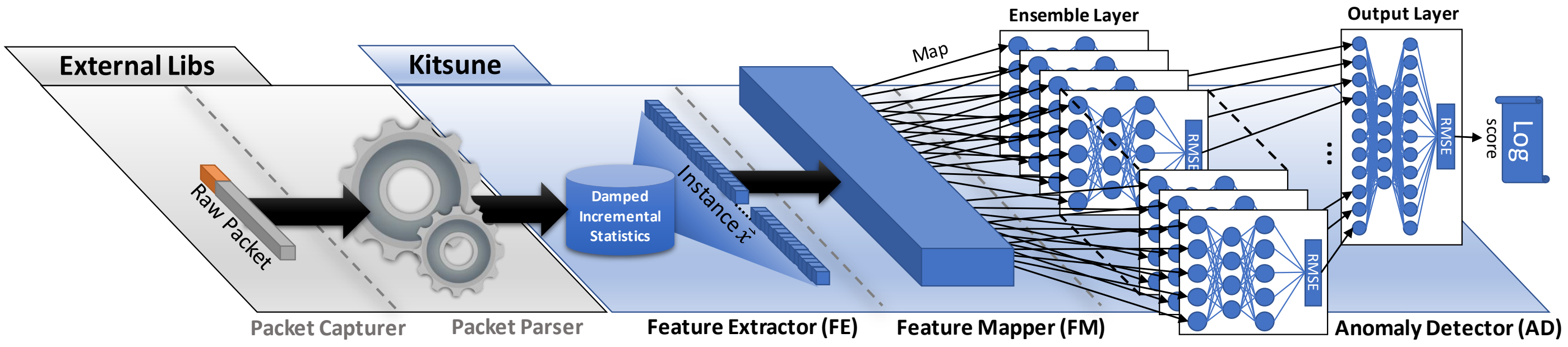

Fig. 3: An illustration of **Kitsune**'s Architecture.

4) The **FM** receives $\vec{x}$...
   - ***Train-mode***: ...and uses $\vec{x}$ to learn a feature map. The map groups features of $\vec{x}$ into sets with a maximum size of $m$ each. Nothing is passed to the AD until the map is complete. At the end of *train-mode*, the map is passed to the AD, and the AD uses the map to build the ensemble architecture (each set forms the inputs to an autoencoder in the ensemble).
   - ***Exec-mode***: ...and the learned mapping is used to create a collection of small instances $\mathbf{v}$ from $\vec{x}$, which is then passed to the respective autoencoders in the ensemble layer of the AD.
5) The **AD** receives $\mathbf{v}$...
   - ***Train-mode***: ..if and uses $\mathbf{v}$ to train the ensemble layer. The RMSE of the *forward-propagation* is then used to train the output layer. The largest RMSE of the output layer is set as $\phi$ and stored for later use.
   - ***Exec-mode***: ...and executes $\mathbf{v}$ across all layers. If the RMSE of the output layer exceeds $\phi\beta$, then an alert is logged with packet details.
6) The original packet, $\vec{x}$, and $\mathbf{v}$ are discarded.

We discuss how the FE, FM, and AD components work in greater detail.

### B. Feature Extractor (FE)

Feature extraction is the process of obtaining or engineering a vector of values which describe a real world observation. In network anomaly detection, it is important to extract features which capture the context and purpose of each packet traversing the network. For example, consider a single TCP SYN packet. The packet may be a benign attempt to establish a connection with a server, or it may be one of millions of similar packets sent in an attempt to cause a denial of service attack (DoS). As another example, consider a video stream sent from an IP surveillance camera. Although the contents of the packets are legitimate, there may suddenly appear a consistently significant rise in jitter. This may indicate the traffic is being sniffed in a man-in-the-middle attack.

These are just some example of attacks where temporal-statistical features could help detect anomalies. The challenge with extracting these kinds of features from network traffic is that (1) packets from different channels (conversations) are interleaved, (2) there can be many channels at any given moment, (3) the packet arrival rate can be very high. The naive approach is to maintain a window of packets from each channel, and to continuously compute statistics over those windows. However, it clear how this can become impractical in terms of memory, and doesn't scale very well.

For this reason, we designed a framework for high speed feature extraction of temporal statistics, over a dynamic number of data streams (network channels). The framework has a small memory footprint since it uses incremental statistics maintained over a damped window. Using a damped window means that the extracted features are temporal (capture the recent behavior of the packet's channel), and that an incremental statistic can be deleted when its dampening weight becomes zero (saving additional memory). The framework has a $O(1)$ complexity because the collection of incremental statistics is maintained in a hash table. The framework also maintains useful 2D statistics which capture the relationship between the rx and tx traffic of a connection.

We will now briefly describe how damped incremental statistics work. Afterwards, we will enumerate the statistical features extracted by the FE to produce the instance $\vec{x}$.

*1) Damped Incremental Statistics:* Let $S = \{x_1, x_2, \ldots\}$ be an unbounded data stream where $x_i \in \mathbb{R}$. For example, $S$ can be a sequence of observed packet sizes. The mean, variance, and standard deviation of $S$ can be updated incrementally by maintaining the tuple $IS := (N, LS, SS)$, where $N$,$LS$, and $SS$ are the number, linear sum, and squared sum of instances seen so far. Concretely, the update procedure for inserting $x_i$ into $IS$ is $IS \leftarrow (N{+}1, LS{+}x_i, SS{+}x_i^2)$, and the statistics at any given time are $\mu_S = \frac{LS}{N}$, $\sigma_S^2 = |\frac{SS}{N} - \left(\frac{LS}{N}\right)^2|$, and $\sigma_S = \sqrt{\sigma_S^2}$.

In order to extract the current behavior of a data stream, we must forget older instances. The naive approach is to maintain a sliding window of values. However, this approach has a memory and runtime complexity of $O(n)$, in contrast to $O(1)$ for an incremental statistic. Furthermore, the sliding window approach does not consider the amount of time spanned by the window. For example, the last 100 instances could have arrived over the last hour or in the last few seconds.

The solution to this is to use damped incremental statistics. In a damped window model, the weight of older values are exponentially decreased over time. Let $d$ be the decay function defined as

$$d_\lambda(t) = 2^{-\lambda t} \tag{6}$$

where $\lambda > 0$ is the decay factor, and $t$ is the time elapsed since the last observation from stream $S_i$. The tuple of a damped incremental statistic is defined as $IS_{i,\lambda} := (w, LS, SS, SR_{ij}, T_{last})$, where $w$ is the current weight, $T_{last}$ is the timestamp of the last update of $IS_{i,\lambda}$, and $SR_{ij}$ is the sum of residual products between streams $i$ and $j$ (used for computing 2D statistics). To update $IS_\lambda$ with $x_{cur}$ at time $t_{cur}$, Algorithm 3 is performed.

TABLE I: Summary of the incremental statistics which can be computed from $S_i$ and $S_j$.

| Type | Statistic | Notation | Calculation |
|---|---|---|---|
| 1D | **Weight** | $w$ | $w$ |
| | **Mean** | $\mu_{S_i}$ | $LS/w$ |
| | **Std.** | $\sigma_{S_i}$ | $\sqrt{\lvert SS/w - (LS/w)^2\rvert}$ |
| 2D | **Magnitude** | $\lVert S_i, S_j \rVert$ | $\sqrt{\mu_{S_i}^2 + \mu_{S_j}^2}$ |
| | **Radius** | $R_{S_i,S_j}$ | $\sqrt{\left(\sigma_{S_i}^2\right)^2 + \left(\sigma_{S_j}^2\right)^2}$ |
| | **Approx. Covariance** | $Cov_{S_i,S_j}$ | $\frac{SR_{ij}}{w_i + w_j}$ |
| | **Correlation Coefficient** | $P_{S_i,S_j}$ | $\frac{Cov_{S_i,S_j}}{\sigma_{S_i}\sigma_{S_j}}$ |

Table I provides a list of the statistics which can be computed from the incremental statistic $IS_{i,\lambda}$. We refer to the statistics whose computation involves one and two incremental statistics as 1D and 2D statistics respectively.

**Algorithm 3:** The algorithm for inserting a new value into a damped incremental statistic.

**procedure:** `update`($IS_{i,\lambda}$,$x_{cur}$,$t_{cur}$,$r_j$)
$\gamma \leftarrow d_\lambda(t_{cur} - t_{last})$ ▷ Compute decay factor
$IS_{i,\lambda} \leftarrow (\gamma w, \gamma LS, \gamma SS, \gamma SR, T_{cur})$ ▷ Process decay
$IS_{i,\lambda} \leftarrow (w{+}1, LS{+}x_{cur}, SS{+}x_i^2, SR_{ij}{+}r_i r_j, T_{cur})$ ▷ Insert value
return $IS_{i,\lambda}$

*2) Features Extracted for Kitsune:* Whenever a packet arrives, we extract a behavioral snapshot of the hosts and protocols which communicated the given packet. The snapshot consists of 115 traffic statistics capturing a small temporal window into: (1) the packet's sender in general, and (2) the traffic between the packet's sender and receiver.

Specifically, the statistics summarize all of the traffic...

- ...originating from this packet's source MAC and IP address (denoted **SrcMAC-IP**).
- ...originating from this packet's source IP (denoted **SrcIP**).
- ...sent between this packet's source and destination IPs (denoted **Channel**).
- ...sent between this packet's source and destination TCP/UDP Socket (denoted **Socket**).

A total of 23 features (capturing the above) can be extracted from a single time window $\lambda$ (see Table II). The FE extracts the same set of features from a total of five time windows: 100ms, 500ms, 1.5sec, 10sec, and 1min into the past ($\lambda = 5, 3, 1, 0.1, 0.01$), thus totaling 115 features.

We note that not every packet applies to every channel type (e.g., there is no socket if the packet does not contain a TCP or UDP datagram). In these cases, these features are zeroed. Thus, the final feature vector $\vec{x}$, which the FE passes to the FM, is always a member of $\mathbb{R}^n$, where $n = 115$.

### *C. Feature Mapper (FM)*

The purpose of the FM is to map $\vec{x}$'s $n$ features (dimensions) into $k$ smaller sub-instances, one sub-instance for each autoencoder in the Ensemble Layer of the AD. Let $\mathbf{v}$ denote the ordered set of $k$ sub-instances, where

$$\mathbf{v} = \{\vec{v_1}, \vec{v_2}, \cdots, \vec{v_k}\} \tag{7}$$

We note that the sub-instances of $\mathbf{v}$ can be viewed as subspaces of $\vec{x}$'s domain $X$.

In order to ensure that the ensemble in the AD operates effectively and with a low complexity, we require that the selected mapping $f(\vec{x}) = \mathbf{v}$:

1) Guarantee that each $\vec{v_i}$ has no more than $m$ features, where $m$ is a user defined parameter of the system. The parameter $m$ affects the collective complexity of the ensemble (see section IV-E).
2) Map each of the $n$ features in $\vec{x}$ exactly once to the features in $\mathbf{v}$. This is to ensure that the ensemble is not too wide.
3) Contain subspaces of $X$ which capture the normal behavior well enough to detect anomalous events occurring in the respective subspaces.
4) Be discovered in a process which is online, so that no more than one at time is stored in memory.

To respect the above requirements, we find the mapping $f$ by incrementally clustering the features (dimensions) of $X$ into $k$ groups which are no larger than $m$. We accomplish this by performing agglomerative hierarchal clustering on incrementally updated summary data.

More precisely, the feature mapping algorithm of the FE performs the following steps:

1) While in train-mode, incrementally update summary statistics with features of instance $\vec{x}$.
2) When train-mode ends, perform hierarchal clustering on the statistics to form $f$.
3) While in execute-mode, perform $f(\vec{x_t}) = \mathbf{v}$, and pass $\mathbf{v}$ to the AD.

In order to ensure that the grouped features capture normal behavior, in the clustering process, we use correlation as the distance measure between two dimensions. In general, the correlation distance $d_{cor}$ between two vectors $u$ and $v$ is defined as

$$d_{cor}(u, v) = 1 - \frac{(u - \bar{u}) \cdot (v - \bar{v})}{\lVert(u - \bar{u})\rVert_2 \lVert(v - \bar{v})\rVert_2} \tag{8}$$

where $\bar{u}$ is the mean of the elements in vector $u$, and $u \cdot v$ is the dot product.

We will now explain how the correlation distances between features can be summarized incrementally for the purpose of clustering. Let $n_t$ be the number of instances seen so far. Let $\vec{c}$ be an $n$ dimensional vector containing the linear sum of each feature's values, such that element $c^{(i)} = \sum_{t=0}^{n_t} x_t^{(i)}$ for feature $i$ at the time index $t$. Similarly, Let $\vec{c}_{\mathrm{r}}$ denote a vector containing the summed residuals of each feature, such that $c_r^{(i)} = \sum_{t=0}^{n_t} \left( x_t^{(i)} - \frac{c^{(i)}}{n_t} \right)$. Similarity, Let $\vec{c}_{\mathrm{rs}}$ denote a vector containing the summed squared residuals of each feature, such that $c_{rs}^{(i)} = \sum_{t=0}^{n_t} \left( x_t^{(i)} - \frac{c^{(i)}}{n_t} \right)^2$. Let $C$ be the $n$-by-$n$ partial correlation matrix, where

$$[C_{i,j}] = \sum_{t=0}^{n_t} \left( \left( x_t^{(i)} - \frac{c^{(i)}}{n_t} \right) \left( x_t^{(j)} - \frac{c^{(j)}}{n_t} \right) \right) \qquad (9)$$

is the sum of products between the residuals of features $i$ and $j$. Let $D$ be the correlation distance matrix between each features of $X$.

TABLE II: The statistics (features) extracted from each time window $\lambda$ when a packet arrives.

| The packet's... | Statistics | Aggregated by | # Features | Description of the Statistics |
|---|---|---|---|---|
| **...size** | $\mu_i, \sigma_i$ | SrcMAC-IP, SrcIP, Channel, Socket | 8 | *Bandwidth of the outbound traffic* |
| **...size** | $\|S_i, S_j\|, R_{S_i,S_j}, Cov_{S_i,S_j}, P_{S_i,S_j}$ | Channel, Socket | 8 | *Bandwidth of the outbound and inbound traffic together* |
| **...count** | $w_i$ | SrcMAC-IP, SrcIP, Channel, Socket | 4 | *Packet rate of the outbound traffic* |
| **...jitter** | $w_i, \mu_i, \sigma_i$ | Channel | 3 | *Inter-packet delays of the outbound traffic* |

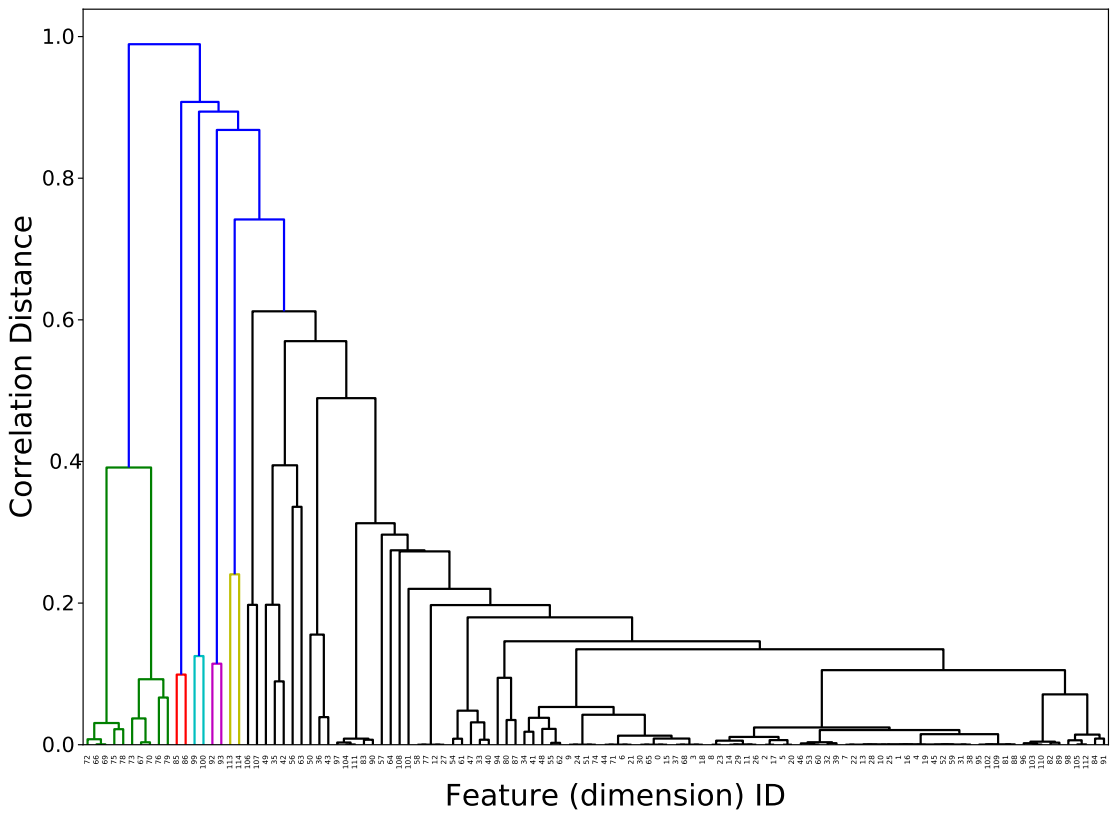

Fig. 4: An example dendrogram of the 115 features clustered together, from one million network packets.

Using $C$ and $\vec{c}_{rs}$, the correlation distance matrix $D$ can be computed at any time by

$$D = [D_{i,j}] = 1 - \frac{C_{i,j}}{\sqrt{c_{rs}^{(i)}} \sqrt{c_{rs}^{(j)}}} \qquad (10)$$

Now that we know how to obtain the distance matrix $D$ incrementally, we can perform agglomerative hierarchal clustering on $D$ to find $f$. Briefly, the algorithm starts with $n$ clusters, one cluster for each point represented by $D$. It then searches for the two closest points and joins their associated clusters. This search and join procedure repeats until there is one large cluster containing all $n$ points. The tree which represents the discovered links is called a dendrogram (pictured in Fig. 4). For further information on the clustering algorithm, we refer the reader to [34]. Typically, hierarchal clustering cannot be performed on large datasets due to its complexity. However, our distance matrix is small ($n$ being in the order of a few hundred) and therefore practical to compute on-site.

Finally, with the dendrogram of $D$, we can easily find $k$ clusters (groups of features) where no cluster is larger than $m$. The procedure is to break the dendrogram's largest link (i.e., the top most node) and then check if all found clusters have a size less then $m$. If there is at least one cluster with a size greater than $m$, then we repeat the process on the exceeding clusters. At the end of the procedure, we will have $k$ groups of features with a strong inter-correlation, where no single group is larger than $m$. These groupings are saved, and used to perform the mapping $f$.

The algorithm described in this section is suitable for online and on-site processing because (1) it never stores more than one instance in memory, (2) it uses very little memory ($On^2$ during train-mode) and, (3) is very fast since the update procedures require updating small $n$-by-$n$ distance matrix.

### D. Anomaly Detector (AD)

As depicted in Fig. 3, the AD component contains a special neural network we refer to as a *KitNET* (**Kit**sune **NET**work). *KitNET* is an unsupervised ANN designed for the task of online anomaly detection. *KitNET* is composed of two layers of autoencoders: the Ensemble Layer and the Output Layer.

**Ensemble Layer:** An ordered set of $k$ three-layer autoencoders, mapped to the respective instances in $\mathbf{v}$. This layer is responsible for measuring the independent abnormality of each subspace (instance) in $\mathbf{v}$. During *train-mode*, the autoencoders learn the normal behavior of their respective subspaces. During both *train-mode* and *execute-mode*, each autoencoder reports its RMSE reconstruction error to the Output Layer.

**Output Layer:** A three-layer autoencoder which learns the normal (i.e., *train-mode*) RMSEs of the Ensemble Layer. This layer is responsible for producing a final anomaly score, considering (1) the relationship between subspace abnormalities, and (2) naturally occurring noise in the network traffic.

We will now detail how *KitNET* operates the Ensemble and Output Layers.

*1) Initialization:* When the AD receives the first set of mapped instances $\mathbf{v}$ from the FM, the AD initializes *KitNET*'s architecture using $\mathbf{v}$ as a blueprint. Concretely, let $\theta$ denote an entire autoencoder, and let $L^{(1)}$ and $L^{(2)}$ denote the Ensemble and Output Layers respectively. $L^{(1)}$ is defined as the ordered set

$$L^{(1)} = \{\theta_1, \theta_2, \dots, \theta_k\} \qquad (11)$$

such that autoencoder $\theta_i \in L^{(1)}$ has three layers of neurons: $\mathbf{dim}(\vec{v}_i)$ neurons in the input and output, and $\lceil \beta \cdot \mathbf{dim}(\vec{v}_i) \rceil$

**Algorithm 4:** The *back-propagation* training algorithm for *KitNET*.

**procedure:** $\texttt{train}(L^{(1)}, L^{(2)}, \mathbf{v})$
`// Train Ensemble Layer`
1 $\vec{z} \leftarrow \text{zeros}(k)$ ▷ *init input for* $L^{(2)}$
2 **for** *(*$\theta_i$ *in* $L^{(1)}$*)* **do**
3 $\quad \vec{v_i}' = \texttt{norm}_{0-1}(\vec{v_i})$
4 $\quad A_i, \vec{y_i} \leftarrow h_{\theta_i}(\vec{v_i}')$ ▷ *forward propagation*
5 $\quad deltas_i \leftarrow b_{\theta_i}(\vec{v_i}', \vec{y_i})$ ▷ *backward propagation*
6 $\quad \theta_i \leftarrow \texttt{GD}_\ell(A_i, deltas_i)$ ▷ *weight update*
7 $\quad \vec{z}[i] \leftarrow \texttt{RMSE}(\vec{v_i}', \vec{y_i})$ ▷ *set error signal*
8 **end**
`// Train Output Layer`
9 $\vec{z}' = \texttt{norm}_{0-1}(\vec{z})$ $A_0, \vec{y_0} \leftarrow h_{\theta_0}(\vec{z}')$ ▷ *forward propagation*
10 $deltas_0 \leftarrow b_{\theta_0}(\vec{z}', \vec{y_0})$ ▷ *backward propagation*
11 $\theta_0 \leftarrow \texttt{GD}_\ell(A_0, deltas_0)$ ▷ *weight update*
12 return $L^{(1)}, L^{(2)}$

neurons in the inner layer, where $\beta \in (0, 1]$ (in our experiments we take $\beta = \frac{3}{4}$). Fig. 5 illustrates the described mapping between $\vec{v_i} \in \mathbf{v}$ and $\vec{\theta_i} \in L^{(1)}$.

$L^{(2)}$ is defined as the single autoencoder $\theta_0$, which has $k$ input and output neurons, and $\lceil k \cdot \beta \rceil$ inner neurons.

Layers $L^{(1)}$ and $L^{(2)}$ are not connected via the weighted synapses found the common ANN. Rather, the inputs to $L^{(2)}$ are the 0-1 normalized RMSE error signals from each respective autoencoder in $L^{(1)}$. Signaling the aggregated errors (RMSEs) of each autoencoder in $L^{(1)}$, as opposed to signaling from each individual neuron of $L^{(1)}$, reduces the complexity of the network. Finally, the weights of autoencoder $\theta_i$ in *KitNET* is initialized with random values from the uniform distribution $\mathcal{U}(\frac{-1}{\mathbf{dim}(\vec{v_i})}, \frac{1}{\mathbf{dim}(\vec{v_i})})$.

*2) Train-mode:* Training *KitNET* is slightly different than training a common ANN network, as described in section III. This is because *KitNET* signals RMSE reconstruction errors between the two main Layers of the network. Furthermore, *KitNET* is trained using SGD using each observed instance $\mathbf{v}$ exactly once. The algorithm for training *KitNET* on a single instance is presented in Algorithm 4.

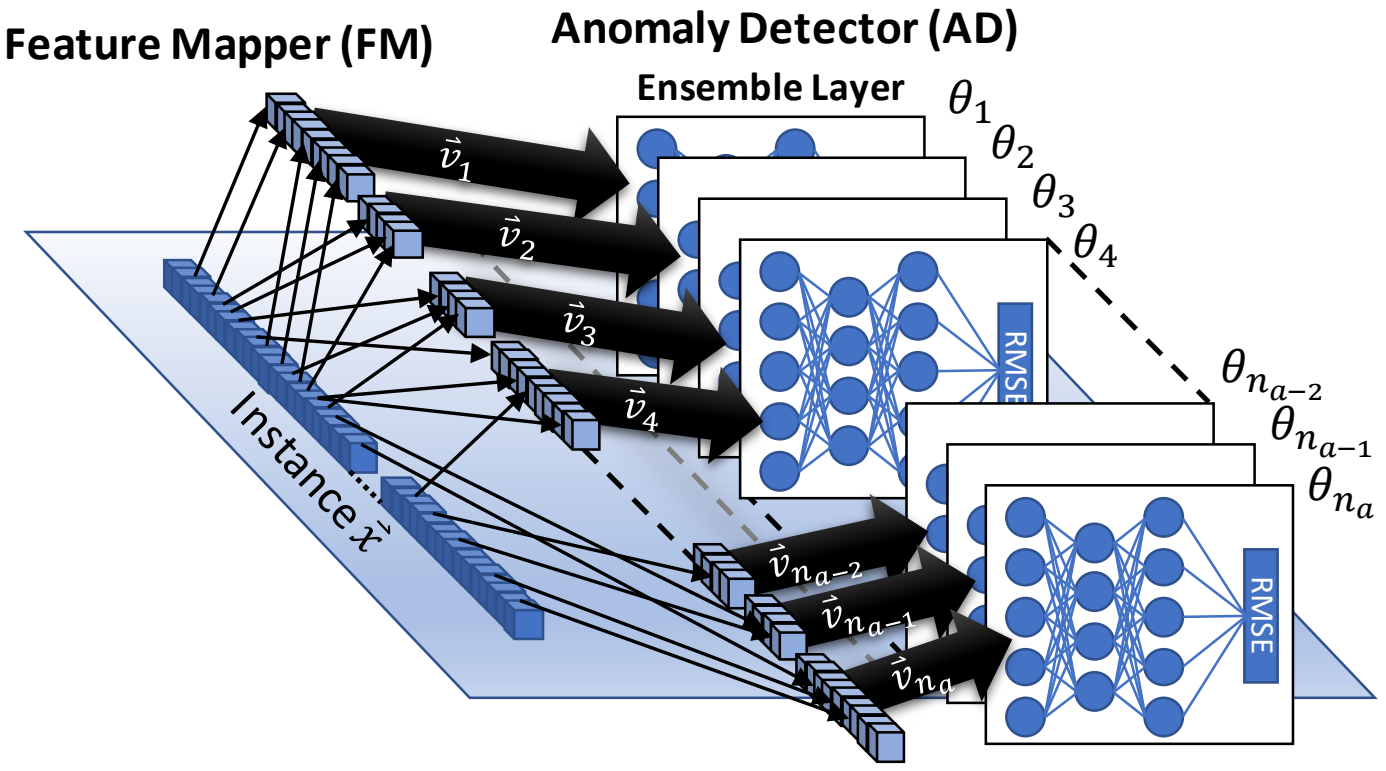

Fig. 5: An illustration of the mapping process between the FM and the Ensemble Layer of *KitNET*: a sub-instance $\vec{v_i}$ is mapped from $\vec{x}$, which is then sent to the autoencoder $\theta_i$.

**Algorithm 5:** The execution algorithm for *KitNET*.

**procedure:** $\texttt{execute}(L^{(1)}, L^{(2)}, \mathbf{v})$
`// Execute Ensemble Layer`
1 $\vec{z} \leftarrow \text{zeros}(k)$ ▷ *init input for* $L^{(2)}$
2 **for** *(*$\theta_i$ *in* $L^{(1)}$*)* **do**
3 $\quad \vec{v_i}' = \texttt{norm}_{0-1}(\vec{v_i})$
4 $\quad A_i, \vec{y_i} \leftarrow h_{\theta_i}(\vec{v_i}')$ ▷ *forward propagation*
5 $\quad \vec{z}[i] \leftarrow \texttt{RMSE}(\vec{v_i}', \vec{y_i})$ ▷ *set error signal*
6 **end**
`// Execute Output Layer`
7 $\vec{z}' = \texttt{norm}_{0-1}(\vec{z})$
8 $A_0, \vec{y_0} \leftarrow h_{\theta_0}(\vec{z_1}')$ ▷ *forward propagation*
9 return $\leftarrow \texttt{RMSE}(\vec{z}', \vec{y_0})$

We note that in order to perform the 0-1 normalization on line 3, each autoencoder must maintain a record of the largest and smallest value seen for each input feature. Furthermore, these maximum and minimum records are only updated during *train-mode*.

Similar to the discussion in section III-E, *KitNET* must be trained on normal data (without the presence of attacks). This is a common assumption [35], [36] and is practical in many types of computer networks, such as IP camera surveillance systems. Furthermore, there are methods for filtering the training data in order to reduce the impact of the possible preexisting attacks in the network [37], [38].

*3) Execute-mode:* In *execute-mode*, *KitNET* does not update any of its internal parameters. Instead, *KitNET* performs forward propagation through the entire network, and returns $L^{(2)}$'s RMSE reconstruction error. The execution procedure of *KitNET* is presented in Algorithm 5.

$L^{(2)}$'s reconstruction error measures the instance's abnormality with respect to the relationships between the subspaces in $\mathbf{v}$. For example, consider two autoencoders from the Ensemble Layer $\theta_i, \theta_j \in L^{(1)}$. If the of RMSE $\theta_i$ and $\theta_j$ correlate during *train-mode*, then a lack of correlation in *execute-mode* may be considered a more significant anomaly than say the simple sum of their independent RMSEs (similarly in vice versa). Since the Output Layer $L^{(2)}$ learns these relationships (and other complex relationships) during *train-mode*, $L^{(2)}$'s reconstruction error of $L^{(1)}$'s RMSEs will reflect these anomalies.

*4) Anomaly Scoring:* The output of *KitNET* is the RMSE anomaly score $s \in [0, \infty)]$, as described in section III-E. The larger the score $s$, the greater the anomaly. To use $s$, one must determine an anomaly score cutoff threshold $\phi$. The naive approach is to set $\phi$ to the largest score seen during *train-mode*, where we assume that all instances represent normal traffic. Another approach is to select $\phi$ probabilistically. Concretely, one may (1) fit the outputted RMSE scores to log-normal or non-standard distribution, and then (2) raise an alert if $s$ has a very low probability of occurring. A user of *KitNET* should decide the best method of selecting $\phi$ according to his/her application of the algorithm. In section V, we evaluate **Kitsune**'s detection capabilities based on its raw RMSE scores.

### *E. Complexity*

As a baseline, we shall compare the complexity of *KitNET* to a single three-layer autoencoder over the same feature space $\vec{x} \in \mathbb{R}^n$, with the compression layer ratio $\beta \in (0, 1]$.

The complexity of executing the single autoencoder is as follows: In section III-F, we found that the complexity of activating layer $l^{(i+1)}$ in an ANN is $O\left(l^{(i)} \cdot l^{(i+1)}\right)$. Therefore, the complexity of executing the single autoencoder is

$$O(n \cdot \beta n + \beta n \cdot n) = O(n^2) \tag{12}$$

The complexity of executing *KitNET* is as follows: We remind the reader that $k$ denotes the number of subspaces (autoencoders) selected by the FM, and that $m$ is an input parameter of the system which defines the maximum number of inputs for any given autoencoder in $L^{(1)}$. The complexity of executing $L^{(1)}$ and $L^{(2)}$ are $O(km^2)$ and $O(k^2)$ respectively. Since the variable $m \in 1, 2, \ldots, 10$ is a constant parameter of the system, the total complexity of *KitNET* is

$$O(km^2 + k^2) = O(k^2) \tag{13}$$

The result of (13) tells us that the complexity of the Ensemble Layer scales linearly with $n$, but the complexity of the Output Layer depends on how many autoencoders (subspaces) are designated by the FM. Concretely, the best case scenario is where $k = \frac{n}{m}$, in which case performance increased by a factor of $m$. The worst case scenario is where the FM designates nearly every single feature its own autoencoder in $L^{(1)}$. In this case, $k = n$, and *KitNET* operates as a single wide autoencoder meaning that there is no performance gain. This will also occur if the user sets $m = 1$ or $m = n$.

However, it is very rare for the latter case to occur on a natural dataset. This is because it would mean that the dendrogram from the FM's clustering process in section IV-C is completely imbalanced. For example,

$$d_{cor}(x^{(1)}, x^{(2)}) < d_{cor}(x^{(2)}, x^{(3)}) < d_{cor}(x^{(3)}, x^{(4)}) < \ldots \tag{14}$$

where $x^i$ is the $i$-th dimension of $\mathbb{R}^n$. Therefore, it can be expected that in the presence of many features, *KitNET* will have a runtime which is faster than a single autoencoder or stacked autoencoder. Finally, the complexity of training *KitNET* is also $O(k^2)$ since we learn from each instance only once (see Algorithm 4). This can be contrasted to ANN-based classifiers which typically make multiple passes (epochs) over the training set.

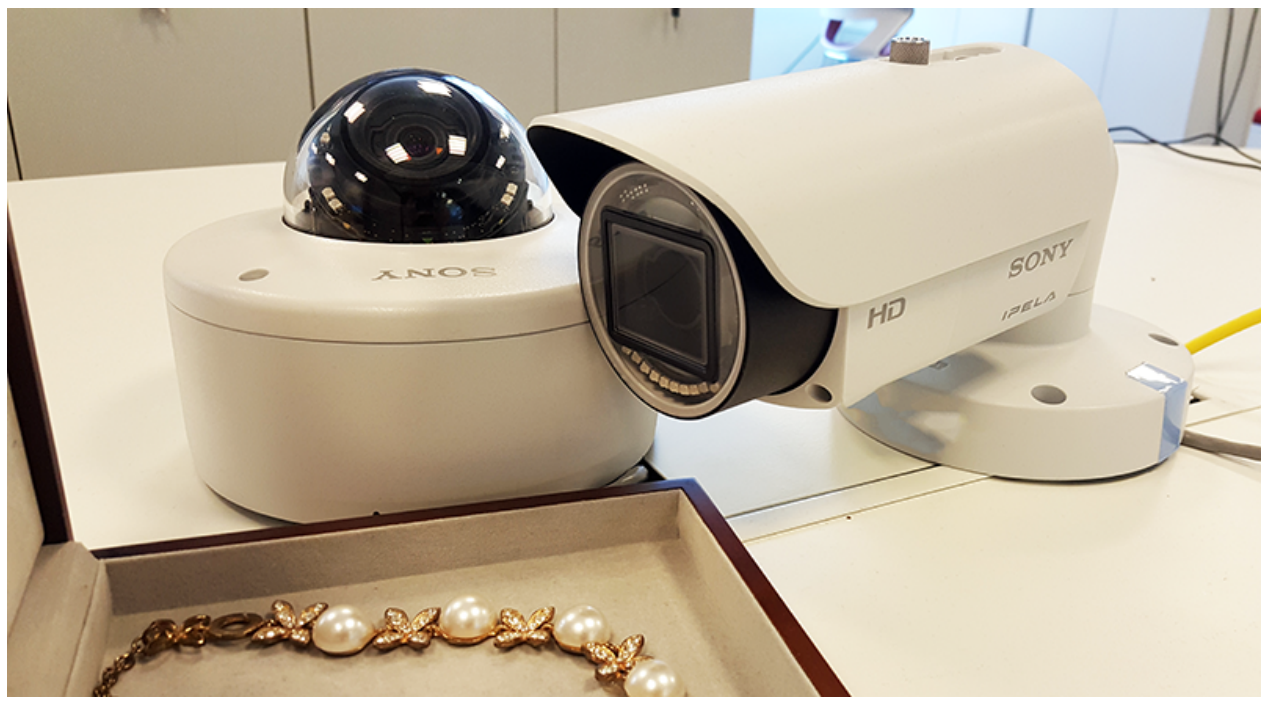

Fig. 6: Two of the cameras used in the IP camera video surveillance network. Left: *SNC-EM602RC*. Right: *SNC-EB602R*.

## V. Evaluation

In this section, we provide an evaluation of **Kitsune** in terms of its detection and runtime performance. We open by describing the datasets, followed by the experiment setup, and finally, close by presenting our results.

### *A. Datasets*

The goal of Kitsune is to provide a light weight IDS which can handle many packets per second on a simple router. Given this goal, we evaluated **Kitsune**'s capabilities in detecting attacks in a real IP camera video surveillance network. The network (pictured in Fig. 7) consists of two deployments of four HD surveillance cameras each. The cameras in the deployments are powered via PoE, and are connected to the DVR via a site-to-site VPN tunnel. The DVR at the remote site provides users with global accessibility to the video streams via a client-to-site VPN connection. The cameras used in the network, and their configurations are described in Table IV. Fig. 6 pictures two of the eight cameras used in our setup.

There are a number of attacks which can performed on the video surveillance network. However, the most critical attacks affect the availability and integrity of the video uplinks. For example, a SYN flood on a target camera, or a man in the middle attack involving video injection into a live video stream. Therefore, in our evaluation, we focused on these types of attacks. Table III summarizes the attack datasets used in our experiments, and Fig. 7 illustrates the location (vector) of the attacker for each respective attack. The Violation column in Table III indicates the attacker's security violation on the network's confidentiality (C), integrity (I), and availability (A). All datasets where recorded from the *packet capture point* as indicated in Fig. 7.

To setup the active wiretap, we used a Raspberry PI 3B as a physical network bridge. The PI was given a USB-to-Ethernet adapter to provide the second Ethernet port, and then placed physically in the middle of the cable.

We note that for some of the attacks, the malicious packets did not explicitly traverse the router on which **Kitsune** is connected to. In these cases, the FE components implicitly captures these attacks as a result of statistical changes in the network's behavior. For example, the man in the middle attacks affect the timing of the packets, but not necessarily the contents of the packets themselves.

In order to evaluate **Kitsune** on a nosier network, we used an additional network. The additional network was a Wi-Fi network populated with 9 IoT devices, and three PCs. The IoTs were a thermostat, baby monitor, webcam, two different doorbells, and four different cheap security cameras. On this particular network, we infected a one of the security cameras with a real sample of the Mirai botnet malware.

### *B. Experiment Setup*

Offline algorithms typically perform better than online algorithms. This is because offline algorithms have access to the entire dataset during training and can perform multiple passes over the data. However, online algorithms are useful when resources, such as the computational power and memory, are limited. In our evaluations, we compare **Kitsune** to both online and offline algorithms. The online algorithms provide a baseline for Kitsune as an online anomaly detector, and the offline algorithms provide a perspective on the Kitsune's performance as a sort of upperbound. As an additional baseline, we evaluate Suricata [39] –a signature-based NIDS. Suricata is an open source NIDS which is similar to the Snort NIDS, but is parallelized over multiple threads. Signature-based NIDS have a much lower false positive rate than anomaly-based NIDS. However, they are incapable of detecting unknown threats or abnormal/abstract behaviors. We configured Suricata to use the 13,465 rules from the Emerging Threats repository [40].

TABLE III: The datasets used to evaluate **Kitsune**.

| Attack Type | Attack Name | Tool | **Description:** *The attacker...* | Violation | Vector | # Packets | Train [min.] | Execute [min.] |
|---|---|---|---|---|---|---|---|---|
| *Recon.* | **OS Scan** | Nmap | *...scans the network for hosts, and their operating systems, to reveal possible vulnerabilities.* | **C** | 1 | 1,697,851 | 33.3 | 18.9 |
| | **Fuzzing** | SFuzz | *...searches for vulnerabilities in the camera's web servers by sending random commands to their cgis.* | **C** | 3 | 2,244,139 | 33.3 | 52.2 |
| *Man in the Middle* | **Video Injection** | Video Jack | *...injects a recorded video clip into a live video stream.* | **C, I** | 1 | 2,472,401 | 14.2 | 19.2 |
| | **ARP MitM** | Ettercap | *...intercepts all LAN traffic via an ARP poisoning attack.* | **C** | 1 | 2,504,267 | 8.05 | 20.1 |
| | **Active Wiretap** | Raspberry PI 3B | *...intercepts all LAN traffic via active wiretap (network bridge) covertly installed on an exposed cable.* | **C** | 2 | 4,554,925 | 20.8 | 74.8 |
| *Denial of Service* | **SSDP Flood** | Saddam | *...overloads the DVR by causing cameras to spam the server with UPnP advertisements.* | **A** | 1 | 4,077,266 | 14.4 | 26.4 |
| | **SYN DoS** | Hping3 | *...disables a camera's video stream by overloading its web server.* | **A** | 1 | 2,771,276 | 18.7 | 34.1 |
| | **SSL Renegotiation** | THC | *...disables a camera's video stream by sending many SSL renegotiation packets to the camera.* | **A** | 1 | 6,084,492 | 10.7 | 54.9 |
| *Botnet Malware* | **Mirai** | Telnet | *...infects IoT with the Mirai malware by exploiting default credentials, and then scans for new vulnerable victims network.* | **C, I** | X | 764,137 | 52.0 | 66.9 |

TABLE IV: The specifications and statistics of the cameras used in the experiments.

| | SNC-EM602RC | SNC-EM600 | SNC-EB600 | SNC-EB602R |
|---|---|---|---|---|
| **Resolution** | 1280x720 | | | |
| **Codec** | H.264/MPEG4 | | | |
| **Frames/Sec** | 15 | | | |
| **Avg. Packets/Sec** | 195 | 350 | 290 | 320 |
| **Avg. Bandwidth** | 1.8 Mbit/s | 1.4 Mbit/s | 1.8 Mbit/s | 1.8 Mbit/s |
| **Protocol** | RTP | RTP | Https *(TLSv1)* | Http/TCP |

For the offline algorithms, we used Isolation Forests (IF) [41] and Gaussian Mixture Models (GMM) [42]. IF is an ensemble based method of outlier detection, and GMM is a statistical method based on the expectation maximization algorithm.For the online algorithms, we used an incremental GMM from [43], and pcStream2 [44]. pcStream2 is a stream clustering algorithm which detects outliers by measuring the Mahalanobis distance of new instances to known clusters.

For each experiment (dataset) every algorithm was trained on the first million packets, and then executed on the remainder. The duration of the first million packets depends on the packet rate of the network at the time of capture. For example, with the OS Scan dataset, each algorithm was trained on the first one million packets, and then executed on the remaining 697,851 packets. Table III lists the relative train and execute periods for each dataset. Each algorithm received the exact same features from Table II.

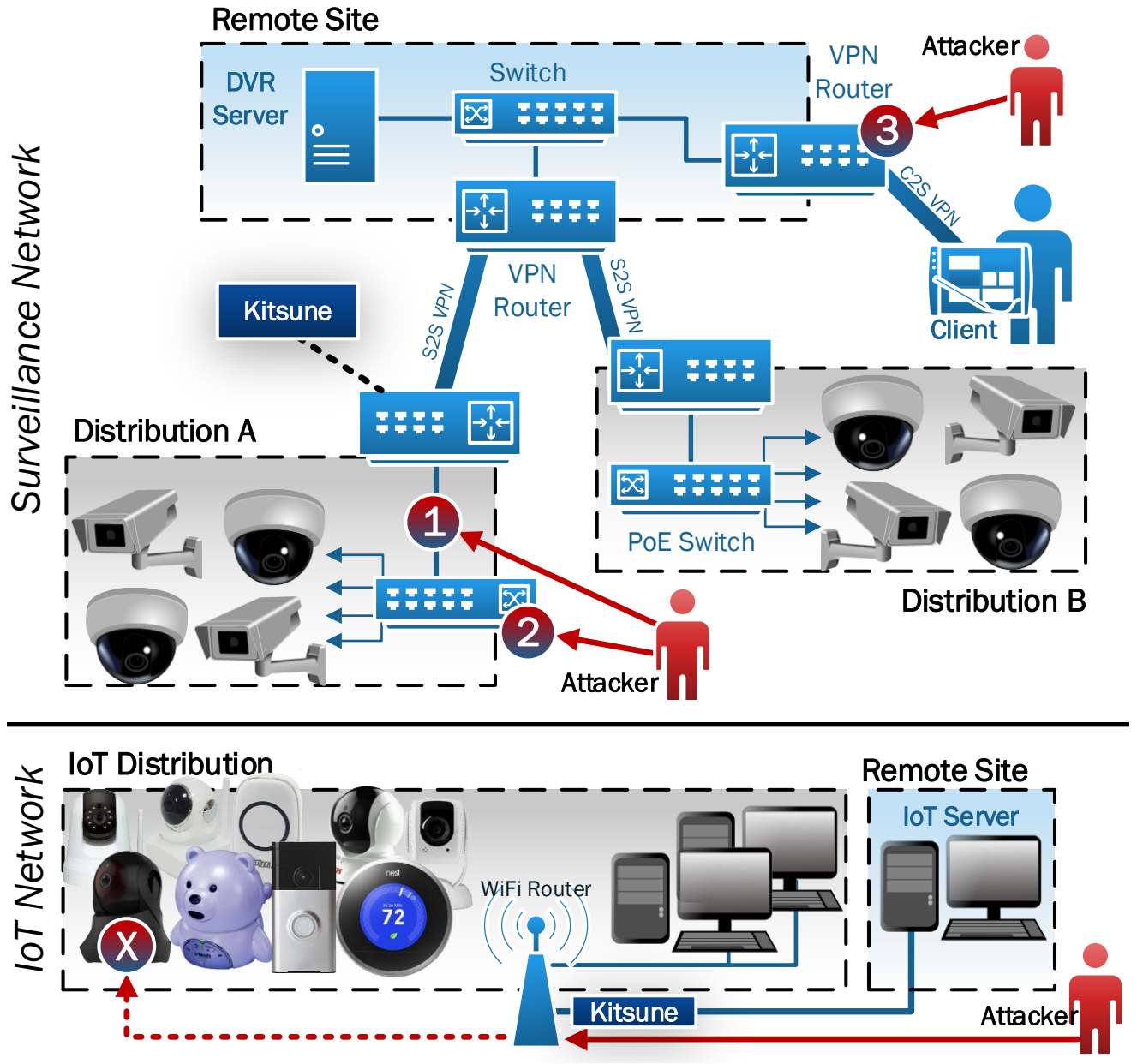

Fig. 7: The network topologies used in the experiments: the surveillance network (top) and the IoT network (bottom).

**Kitsune** has one main parameter, $m \in \{1, 2, \ldots, n\}$, which is the maximum number of inputs for any one autoencoder of *KitNET*'s ensemble. For our detection performance evaluations, we set $m = 1$ and $m = 10$. For all other algorithms, we used the default settings.

### C. Evaluation Metrics

The output of an anomaly detector ($s$) is a value on the range of $[0, \infty)$, where larger values indicate greater anomalies (e.g., the RMSE of an autoencoder). This output is typically normalized such that scores which have a value less than 1 are normal, and greater than 1 are anomalies. The score $s$ is normalized by dividing it by a cutt-off threshold $\phi$. Choosing $\phi$ has a great effect on an algorithm's performance.

The detection performance of an algorithm, on a particular dataset, can be measured in terms of its true positives ($TP$), true negatives ($TN$), false positives ($FP$), and false negatives ($FN$). In our evaluations we measure an algorithm's true positive rate ($TPR = \frac{TP}{TP+FN}$) and the false negative rate ($FNR = \frac{FN}{FN+TP}$) when $\phi$ is selected so that the false positive rate ($FPR = \frac{FP}{FP+TN}$) is very low (i.e., 0.001). We also count the number of true positives where the FPR is zero. These measures capture the amount of malicious packets which were detected correctly, and accidentally missed with few and no false alarms. In network intrusion detection, is it important that there be a minimal number of false alarms since it is time consuming and expensive an analyst to investigate each alarm.

Fig. 8 plots *KitNET*'s anomaly scores before and during a fuzzing attack. The lower blue line is the lowest threshold we might select during the training phase which produces no FPs. The upper blue line is the lowest threshold possible during the attack which produces no FPs (kind of like a global optimum). Another way of looking at these two thresholds is like a best-case and worst case scenarios for threshold selection. Therefore, by measuring the performance of each at these two thresholds, we can get a better idea of an algorithm's potential as an anomaly detector.

To measure the general performance (i.e. with every possible $\phi$), we used the area under the receiver operating characteristic curve (AUC), and the equal error rate (EER). In our context, the AUC is the probability that a classifier will rank a randomly chosen anomalous instance higher than a randomly chosen normal instance. In other words, an algorithm with an AUC of 1 is a perfect anomaly detector on the given dataset, whereas an algorithm with an AUC of 0.5 is randomly guessing labels. The EER is a measure which captures an algorithm's trade-off between its $FNR$ and $FPR$. It is computed as the value of $FNR$ and $FPR$ when they are minimal and equal to one another.

### D. Detection Performance

Figure 9 presents TPR and FNR of each algorithm over each dataset when the threshold is selected so that the FPR is 0 and 0.001. The figure also presents the AUC and EER as well. We remind the reader that the GMM and Isolation Forest are a batch (offline) algorithms, which have full access to the entire dataset and perform many iterations over the dataset. Therefore, these algorithms serve as an optimum goal for us to achieve. In Fig. 9 we see that **Kitsune** performed very well relative to these algorithms. In particular, **Kitsune** performed even better than the GMM in detecting the active wiretap. Moreover, our algorithm achieved a better EER than the GMM on the AR, Fuzzing, Mirai, SSL R., SYN and active wiretap datasets.

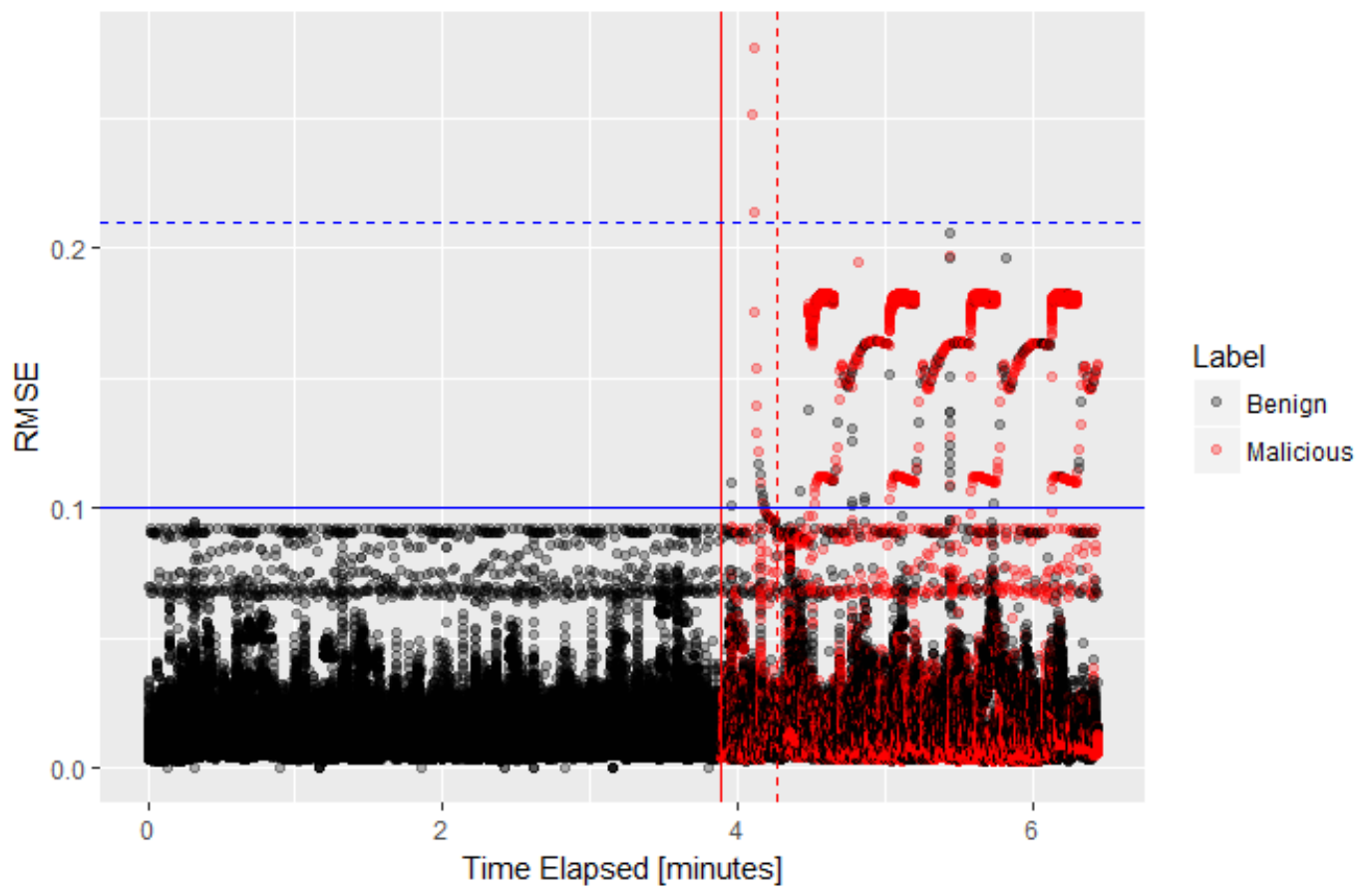

Fig. 8: *KitNET*'s RMSE anomaly scores before and during a fuzzing attack. The red lines indicate when the attacker connects to the network (left) and initiates the attack (right).

As evident from Fig. 9, there is a trade-off between the detection performance and $m$ (runtime performance). Users who prefer better detection performance over speed should use an $m$ which is close to 1 or $n$. Whereas users who prefer speed should use a moderate sized $m$. The **Kitsune** gives the user the ability to adjust this parameter according to the requirements of the user's system. The affect of $m$ on the runtime performance is presented in section V-E.

As a baseline comparison to the performance of online algorithms we compared **Kitsune** to the incremental GMM and pcStream2. Overall, it is clear that **Kitsune** out performs both algorithms in terms of AUC and EER.

The top row of figure 9 presents the maximum number of true positives each algorithm was able to obtain, when the threshold was set so that here were no false positives (e.g., the blue dashed bar in Fig. 8). In other words, these figures show how well each anomaly detector is able to raise the anomaly score of malicious packets above the noise floor. The figures shows that **Kitsune** detects attacks across the datasets better than the other algorithms, and more so than the GMM in most cases. We note that **Kitsune** with $m = 10$ sometimes performs better than $m = 1$. This is because the Output Layer autoencoder lowers the noise floor of the ensemble. This affect can amplify the scores of some outliers.

### E. Runtime Performance

One of **Kitsune**'s greatest advantages is its runtime performance. As discussed in section IV-E, *KitNET*'s ensemble of small autoencoders is more efficient than using a single autoencoder. This is because the ensemble reduces the overall number of operations required to process each instance.

To demonstrate this effect, we performed benchmarks on a Raspberry PI 3B and an Ubuntu Linux VM running on a Windows 10 PC (full details are available in Table V). The experiments were coded in C++, involved $n = 198$ statistical packet features, and were executed on a single core (physical core on the PI and logical core on the Ubuntu VM).

Fig. 10 plots the affect which *KitNET*'s ensemble size has on the packet processing rate. With a single autoencoder in $L^{(1)}$, the PI and PC can handle approx. 1,000 and 7,500 packets per second respectively. However, with 35 autoencoders in $L^{(1)}$, the performance of both environments improve by a factor of 5 to approx. 5400 and 37,300 respectively. Fig. 11 provides a closer look at the PI's packet processing times with $k = 1$ and $k = 35$. This figure shows that using an ensemble can also reduce the variance in the processing times. This may be beneficial in applications where jitter in network traffic is undesirable.

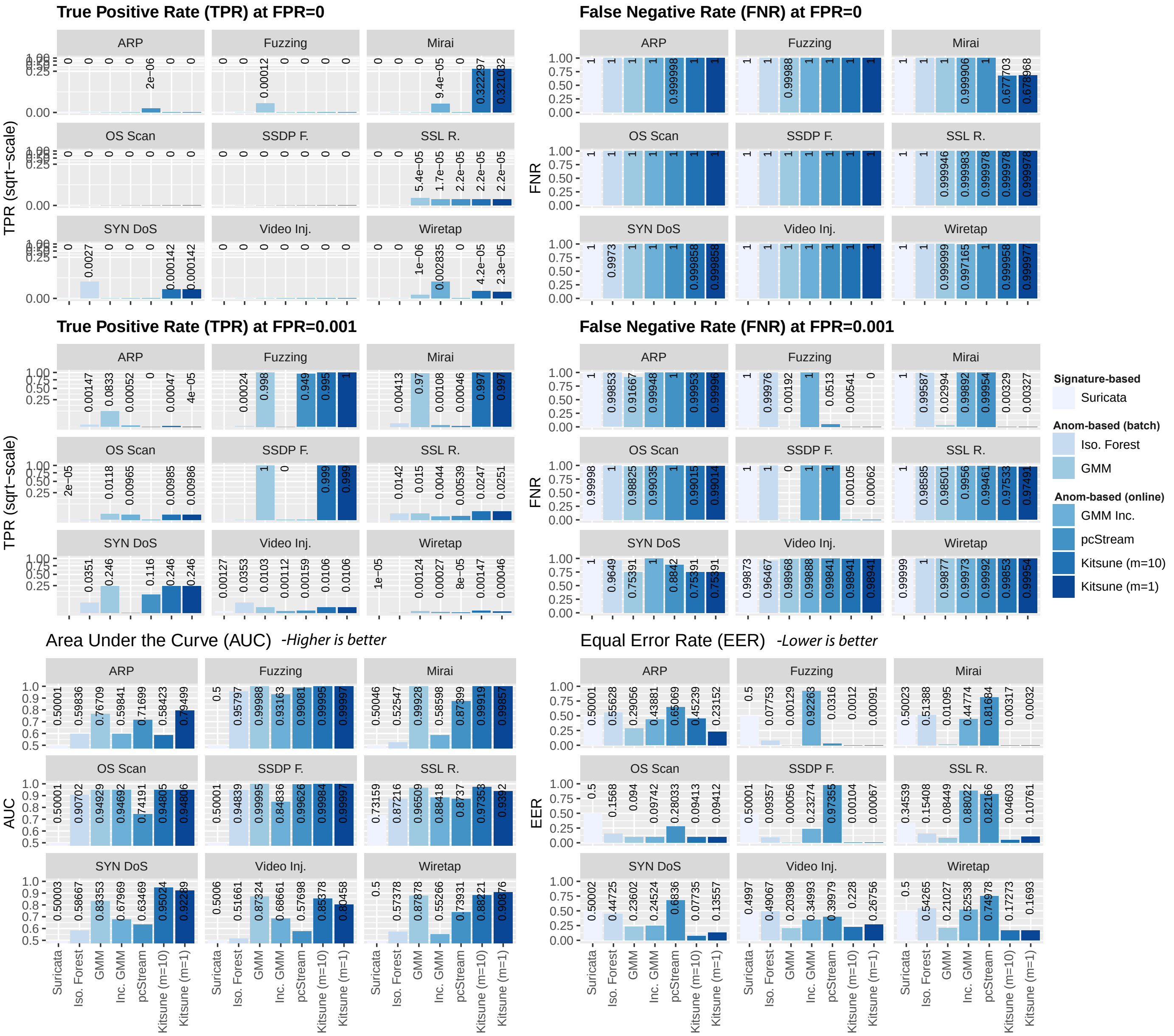

Fig. 9: The experimental results for all algorithms on each of the datasets: the TPR when the FPR is equal to 0.001 (top-left), the FNR when the FPR is equal to 0.001 (top-right), the AUC (bottom-left), and the EER (bottom-right).

The results of the Raspberry PI's benchmark show that a simple network router, with limited resources, can support Kitsune as an NIDS. This means that **Kitsune** is an inexpensive and reliable distributed NIDS solution. We note that since the experiments were run on a single core, there is potential to increase the packet processing rates further. To achieve this, we plan on parallelizing *KitNET* over multiple cores.

## VI. Adversarial Attacks & Countermeasures

When using **Kitsune**, there are several aspects one should consider. First off all, an advanced adversary may attempt to perform adversarial machine learning [45]. When first installed, **Kitsune** assumes that all traffic is benign while in *train-mode*. Therefore, a preexisting adversary may be able to evade **Kitsune**'s detection. However, during *execute-mode*, **Kitsune** will detect new attacks, and new threats as they present themselves. Regardless, a user should be aware of this risk when installing **Kitsune** on a potentially compromised network. As a future work, it would be interesting to find a mechanism which can safely filter out potentially contaminated instances during the training process. For example, we can first execute and see if there is a high anomaly score. If there is, then we will not train from the instance (since we only want to learn from benign instances). By doing so, we can potentially remain in *train-mode* indefinitely.

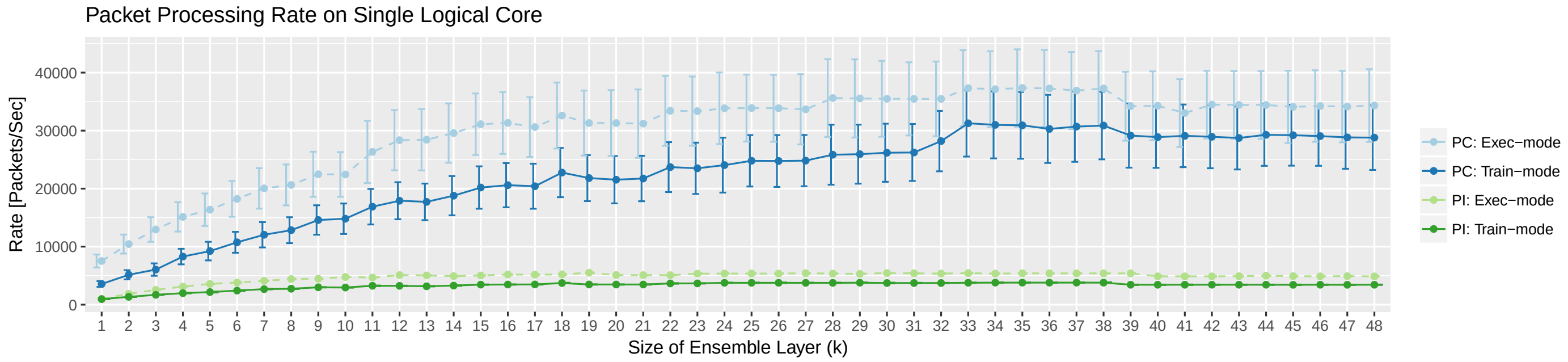

Fig. 10: The affect *KitNET*'s ensemble size ($k$) has on the average packet processing rate, while running on a single core of a Raspberry PI and an Ubuntu VM (PC), using $n = 198$ features.

TABLE V: The environments used to perform the benchmarks.

| | | *Environment 1* **Raspberry PI 3B** | *Environment 2* **Ubuntu VM** |
|---|---|---|---|
| CPU | Type | Broadcom BCM2837 | Intel i7-4790 |
| | Clock | 1.2GHz | 3.60GHz |
| | Cores | 4 | 4 (8 logical) |
| RAM | | 1 GB | 4 GB |

If there is a significant concern that the target network has been contaminated, then one may prefer to use a signature based NIDS, such as Snort. The trade off is that a signature based NIDS cannot automatically detect new or abstract threats (as demonstrated in the evaluation results). A good compromise would be to install an efficient NIDS (such as Snort 3.0 or Suricata) alongside **Kitsune**.

Another threat to **Kitsune** is a DoS attack launched against the FE. In this scenario, an attacker sends many packets with random IP addresses. by doing so, the FE will creates many incremental statistics which eventually consume the device's memory. Although this attack causes a large anomaly, the system may become instable. Therefore, it is highly recommended that the user limit the number of incremental statistics which can be stored in memory. One may note that with a `C++` implementation of **Kitsune**, roughly 1MB of RAM can contain approximately 1,000 network links (assuming five damped windows per link). A good solution to maintaining a small memory footprint is to periodically search and delete incremental statistics with $w_i \approx 0$. In practice, the majority of incremental statistics remain in this state since we use relatively large $\lambda$s (quick decay).

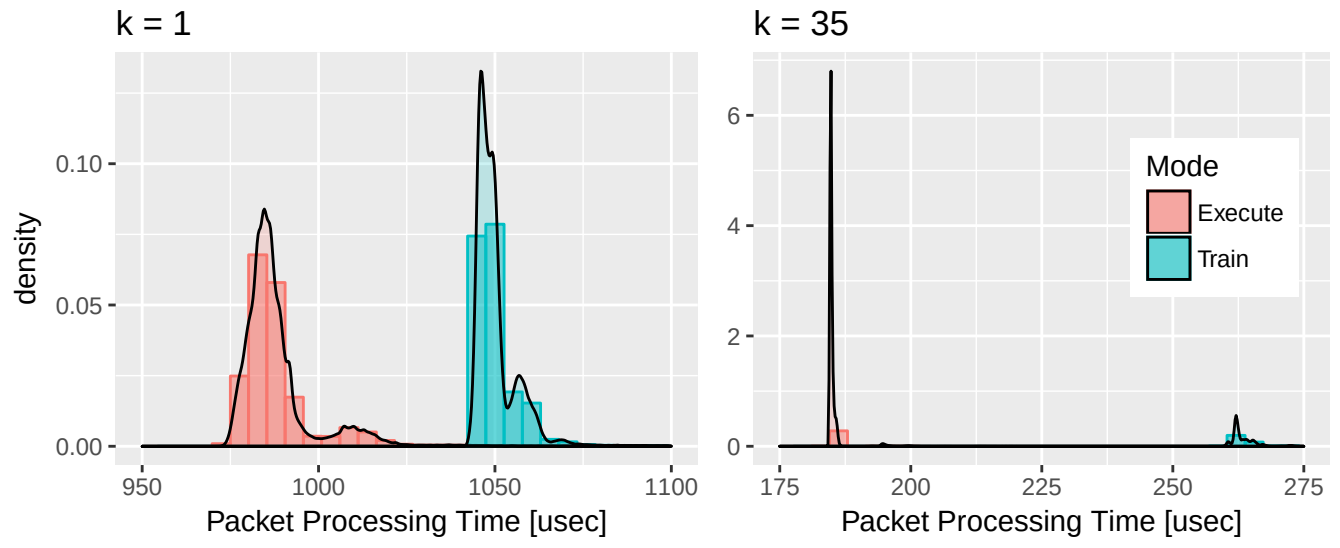

Fig. 11: Density plots of the packet processing times in the PI, with $k = 1$ (top), and $k = 35$ (bottom).

## VII. Conclusion

**Kitsune** is a neural network based NIDS which has been designed to be efficient and plug-and-play. It accomplishes this task by efficiency tracking the behavior of all network channels, and by employing ensemble of auto encoders (*KitNET*) for anomaly detection. In this paper, we discussed the framework's online machine learning process in detail, and evaluated it in terms of detection and runtime performance. *KitNET*, an online algorithm performed nearly as well as other batch / offline algorithms, and in some cases better. Moreover, the algorithm is efficient enough to run on a single core of a Raspberry PI, and has an even greater potential on stronger CPUs.

In summation, there is a great benefit in being able to deploy an intelligent NIDS on simple network devices, especially when the entire deployment process is plug-and-play. We hope that **Kitsune** is beneficial for professionals and researchers alike, and that the *KitNET* algorithm sparks an interest in further developing the domain of online neural network based anomaly detection.

## Acknowledgment

The authors would like to thank Masayuki Nakae, NEC Corporation of America, for his feedback and assistance in building the surveillance camera deployment. The authors would also like to thank Yael Mathov, Michael Bohadana, and Yishai Wiesner for their help in creating the Mirai dataset.